\documentclass[english,aps,manuscript]{emulateapj}
\usepackage[T1]{fontenc}
\usepackage[latin9]{inputenc}
\usepackage{bm}
\usepackage{amsmath}
\usepackage{graphicx}
\usepackage{esint}

\makeatletter

\providecommand{\tabularnewline}{\\}

\@ifundefined{textcolor}{}
{%
 \definecolor{BLACK}{gray}{0}
 \definecolor{WHITE}{gray}{1}
 \definecolor{RED}{rgb}{1,0,0}
 \definecolor{GREEN}{rgb}{0,1,0}
 \definecolor{BLUE}{rgb}{0,0,1}
 \definecolor{CYAN}{cmyk}{1,0,0,0}
 \definecolor{MAGENTA}{cmyk}{0,1,0,0}
 \definecolor{YELLOW}{cmyk}{0,0,1,0}
}

\@ifundefined{date}{}{\date{}}
\date{}
\usepackage{aas_macros}
\usepackage{natbib}

\makeatother

\usepackage{babel}
\begin{document}
\global\long\def\grad{\bm{\nabla}}
\global\long\def\cellavg#1{\langle#1\rangle}
\global\long\def\fc{\mathbf{x}_{h}^{\prime}}
\global\long\def\sc{\mathbf{x}}
\global\long\def\gs{\grad_{\sc}}
\global\long\def\gf{\grad_{\mathbf{h}}}
\global\long\def\gamt{\tilde{\bm{\Gamma}}}
\global\long\def\crctr{\tilde{\mathbf{J}}}

\title{Frequency shifts of resonant modes of the Sun due to near-surface
convective scattering}

\author{J. Bhattacharya, S. Hanasoge, H.M.Antia}

\affiliation{Department of Astronomy and Astrophysics, Tata Institute of Fundamental
Research, Mumbai-400005, India}
\begin{abstract}
Measurements of oscillation frequencies of the Sun and stars can provide
important independent constraints on their internal structure and
dynamics. Seismic models of these oscillations are used to connect
structure and rotation of the star to its resonant frequencies, which
are then compared with observations, the goal being that of minimizing
the difference between the two. Even in the case of the Sun, for which
structure models are highly tuned, observed frequencies show systematic
deviations from modeled frequencies, a phenomenon referred to as the
``surface term''. The dominant source of this systematic effect is
thought to be vigorous near-surface convection, which is not well
accounted for in both stellar modeling and mode-oscillation physics.
Here we bring to bear the method of homogenization, applicable in
the asymptotic limit of large wavelengths (in comparison to the correlation
scale of convection), to characterize the effect of small-scale surface
convection on resonant-mode frequencies in the Sun. We show that the
full oscillation equations, in the presence of temporally stationary
3-D flows, can be reduced to an effective ``quiet-Sun'' wave equation
with altered sound speed, Br\"{u}nt--V\"{a}is\"{a}la frequency
and Lamb frequency. We derive the modified equation and relations
for the appropriate averaging of three dimensional flows and thermal
quantities to obtain the properties of this effective medium. Using
flows obtained from three dimensional numerical simulations of near-surface
convection, we quantify their effect on solar oscillation frequencies,
and find that they are shifted systematically and substantially. We
argue therefore that consistent interpretations of resonant frequencies
must include modifications to the wave equation that effectively capture
the impact of vigorous hydrodynamic convection. 
\end{abstract}
\maketitle

\section{Introduction}

Measurements of oscillation frequencies of stars provide some of the
strongest constraints on stellar structure models. Recent missions
Kepler \citep{borucki_10} and COROT \citep{auvergne_09} have provided
us with high quality asteroseismic data, which have revealed structural
features of Sun-like stars in unprecedented detail. There have been
several successful solar models which have reproduced the oscillation
frequencies to within a few parts in thousand. Despite this degree
of accuracy, the modeled frequencies display systematic deviations
from the observed ones. Understanding the root of these deviations
is important both from theoretical and observational point of view
--- not only is it a window into the missing physics, it is also a
hindrance in fitting stellar models. With the ever increasing precision
in measured frequencies, it is imperative to understand the physical
reason behind the trend and to correct for it while extracting information
about the Sun at the same time. 

There are various reasons why these shifts are observed, which were
categorized as model and modal effects by \citep{rosenthal_99} .
Some of the possible sources for this bias are:
\begin{enumerate}
\item Errors in physics that go into the modeling, including turbulent-pressure
contributions (model effect).
\item Non-adiabaticity prevalent in the outer layers of the Sun (model+modal
effect).
\item Imprecise modeling of propagation of waves through convective flows
near the surface of the Sun (modal effect).
\item Non-linearity in the wave equation (modal effect). 
\end{enumerate}
It is unlikely that any one of these mechanisms would explain the
observed deviations completely, rather a combination of these corrections
would lead to a better estimate of oscillation frequencies, and possibly
eliminate the observed bias. We would like to differentiate between
the first and third points --- the first assumes a hydrodynamic equilibrium
model, but treats oscillations as if they were about an equivalent
hydrostatic background, while the third treats the oscillations about
a dynamic background. In this article we focus on the third aspect,
and propose a method to correct the modal effect. Strong convective
flows with speeds close to the sound speed exist near the surface
of the Sun. We show that advection of traveling waves by these flows
can have a significant impact on the resonant mode frequencies, and
it is therefore important to account for this effect in order to extract
the correct physics from measurements.

Acoustic modes (\textit{p} modes) are standing waves bounded from
below by a lower turning point set by the Lamb frequency, and from
above by an upper turning point set by the acoustic-cutoff frequency.
The upper turning point for modes with frequencies below $2\,\mathrm{mHz}$
is significantly below the surface of the Sun, but modes with frequencies
lying in the range $2\,\mathrm{mHz}$ to $5\,\mathrm{mHz}$ encounter
an upper turning point close to the surface. These are the modes for
which the measured frequencies display a systematic deviation from
model, hence it seems likely that the missing physics lies close to
the surface, and the deviation is often termed `surface effect'. The
deviation is predominantly a function of the frequency of a mode,
hence corrective terms should also be functions of the mode frequency.
An early attempt at mitigating the frequency differences was by \citet{gough90},
whose prescription was in the form of a power law correction based
on the physical process involved: fibrillated magnetic fields lead
to a change in sound speed, and hence an observed frequency $\nu$
would be shifted by $\delta\nu\propto\nu^{3}/\mathcal{I}$, where
$\mathcal{I}$ is the mode inertia; while alternate parametrizations
of convection lead to a shift $\delta\nu\propto1/\left(\nu\mathcal{I}\right)$.
\citet{kjeldsen08} had studied the shifts in the Sun, and with an
empirical power law model $\delta\nu=a\left(\nu/\nu_{0}\right)^{b}$,
were able to correct for the shifts in $\alpha\;\mathrm{Cen}\;\mathrm{A}$,
$\alpha\;\mathrm{Cen}\;\mathrm{B}$ and $\beta\;\mathrm{Hyi}$. Recently
\citet{ball_gizon_14} have shown that a combination of the two effects
studied by Gough --- the cubic shift and the inverse shift --- is
able to fit the deviations better than Kjeldsen's power law. These
corrections are satisfactory from a model-fitting viewpoint, but by
eliminating the frequency differences altogether we lose information
about their origin. 

Despite the high flow speeds, there still exists a small parameter
in the wave equation --- the granulation length scales are usually
much smaller than horizontal wavelengths of low-$\ell$ modes, which
are of the order of the solar radius. \citet{hanasoge_gizon_bal_13}
had designed an alternate approach to deal with small-scale sound
speed perturbation, by expanding the wave equation in terms of the
ratio of the two length scales, followed by averaging over small scales.
They had shown that if the horizontal wavelength of a propagating
wave is much larger than the length scale over which the background
medium fluctuates, it is possible to replace the background by a homogeneous,
coarse-grained one. In this paper, we use a similar analysis in presence
of small-scale flows, and derive an expression for an effective speed.
A way to think of homogenization is as follows: near-surface convection
is a stochastic process, evolving both temporally and spatially. Here
we assume temporally stationarity, with waves propagating through
a frozen configuration of convective flows. Precise measurements of
resonant frequencies require taking spectra of long sequences of raw
observations of the wavefield --- at a minimum, several oscillation
periods but typically much longer than that, on the order of a million
periods, so as to resolve the mode line-width and beat down the noise
sufficiently. Over this time, waves will have propagated through a
large number of realizations of convective flows. Thus one may think
of this as an ensemble of wave equations, each ascribed a set of coefficients
corresponding to different realizations of the process of convection,
drawn from the same probability density function. Each realization
of convective flows corresponds to a wave solution, and the question
then becomes: how to obtain an ensemble average of these wave solutions?
Homogenization provides an answer to this question, specifically in
the limit where the horizontal wavelength is much larger than the
length scale corresponding to the flows. The principle behind homogenization
is that in cases where waves and flows are scale-separated, measurables
such as frequencies can be reproduced by studying wave-propagation
through a quiet, homogeneous medium obtained by appropriately averaging
over small-scale flows. The method has an advantage that the effective
speed which governs wave behaviour on a large-scale emerges naturally
out of the averaging process. In the asymptotic limit, this is the
speed which shall appear in the dispersion relation, and hence shall
determine the observed frequency differences. The method of homogenization
allows an extension to random media where the distribution function
is translationally invariant and bounded \citep{papanicolaou}, and
thus holds promise in progressing beyond our simplified analysis to
real solar models.

Several authors such as \citet{murawski93,duvall98} had studied the
effect of random scatterings by convective flows on surface gravity
waves (\textit{f} modes), arriving at corrections to the dispersion
relation in order to match the observed frequencies and line-widths.
A different treatment to capture the effect of flows on \textit{p}
modes had been carried out by \citet{brown84}, who had proposed a
correction of the form $\left(1-\cellavg{v_{z}^{2}}/c^{2}\right)$
by carrying out a horizontal average over the wave equation. This
form of the correction comes closest to our approach, which generalizes
the analysis to arbitrary small-scale vector flow velocities. An asymptotic
analysis of \textit{p}-mode frequency shifts for small Mach numbers
had been carried out by \citet{stix_zhugshda_04}, but whether the
analysis extends to transonic flows near the solar surface remains
unclear.

\section{Governing Dynamics}

Waves on the Sun are modeled as small-amplitude perturbations about
a background, which we assume to be in a steady state. The background
medium is characterized by its pressure $p\left(\sc\right)$, density
$\rho\left(\sc\right)$, flow velocity ${\bf u}\left(\sc\right)$,
and gravitational potential $\phi\left(\sc\right)$. Waves shall induce
oscillations about the temporarily stationary background, and the
velocity of a moving blob would be given by a sum of the background
flow velocity and the mode velocity. We work with the vector flow
velocity $\mathbf{u}\left(\mathbf{x}\right)$ without splitting it
up into horizontal and vertical components, since both the components
repeat over small scales horizontally and hence can be treated analogously.
In further analysis, we shall suppress explicit spatial and temporal
dependence for notational convenience. The full magnetohydrodynamic
equation for waves traveling through non-uniform flows was analyzed
by \citet{webb05}, we replicate the equation of motion here. The
background flow field $\mathbf{u}$ satisfies the equation 
\begin{equation}
\grad\cdot\left(\rho{\bf uu}+p\mathbf{I}\right)+\rho\grad\phi=\bm{0},\label{eq:back_eom}
\end{equation}
where $\mathbf{I}$ denotes the second-rank identity tensor defined
as $\mathbf{I}_{ij}=\delta_{ij}$. In presence of a flow field ${\bf u}$,
the material derivative of a small-amplitude displacement $\bm{\xi}\left(\sc,t\right)$
is defined as 
\begin{equation}
\dot{\bm{\xi}}=\partial_{t}\bm{\xi}+{\bf u}\cdot\grad\bm{\xi}.\label{eq:material_derivative}
\end{equation}
We assume that there exists a flow field which satisfies the background
equations, and concentrate on the wave equation. The acoustic wave
equation, to linear order in the perturbations, is given by 
\begin{gather}
\partial_{t}\left(\rho\dot{\bm{\xi}}\right)+\grad\cdot\Big\{\rho{\bf u}\dot{\bm{\xi}}+\mathbf{I}\left(p-\rho c^{2}\right)\grad\cdot\bm{\xi}-p\left(\grad\bm{\xi}\right)^{T}\Big\}\nonumber \\
+\rho\bm{\xi}\cdot\grad\grad\phi=\bm{0},\label{eq:full_eqn}
\end{gather}

\noindent where $T$ in the exponent denotes the transpose operator
\citep{webb05}. We shall stick to the Cowling approximation as our
focus is on correcting for the wave speed; the analysis can be extended
to a general case in an analogous manner. 

A propagating wave is scattered out in all direction by the background
flow field, which results in a change in the phase of the forward-traveling
wave. Multiple scattering would lead to repeated phase shifts, leading
to an altered phase velocity of the wave. This modified velocity would
govern the dispersion relation, which would in turn determine the
oscillation frequencies. In order to model this scattering perturbatively,
we resort to the multiple scale approach of spatial homogenization.

\section{Spatial Homogenization}

Low-degree resonant modes on the sun have horizontal wavelengths ($\lambda$)
comparable to the solar radius, which is much larger than the typical
length scales of granulation on the Sun ($L$). In such a case, the
dynamics can be parametrized in terms of the scale ratio $\varepsilon=L/\lambda$.
In the limit $\varepsilon\ll1$, we can replace the rapidly fluctuating
medium by a homogeneous, albeit anisotropic one. To illustrate this
approach, we study a special case of a periodic flow in a Cartesian
box, where we denote the vertical depth by the coordinate $z$, and
label a point on a horizontal plane by the coordinates $\left(x,y\right)$.
The background medium repeats after a length $L$ along both the horizontal
directions $x$ and $y$, and is stratified in the vertical direction.
On a particular horizontal plane, the flow takes the form of tessellated
square cells. To study the dynamics in the asymptotic limit $\varepsilon\ll1$,
we introduce two spatial coordinates which are scale-separated: a
slow coordinate $\sc=\left(x,y,z\right)$ and a horizontal fast coordinate
$\fc=\left(x/\varepsilon,y/\varepsilon\right)$. Note that the vertical
component of mode wavelength can be comparable to the scale of flows,
so the advection terms might not be scale-separable vertically. Bearing
this in mind, scale separation is being carried out strictly in the
horizontal directions, labeled by $x-y$ or equivalently by $\fc$.
The depth coordinate $z$ is common to the flows and the wave, and
it characterizes the dynamics completely in the vertical direction.\textbf{
}In terms of these new coordinates, the spatial gradient can be rewritten
as 
\begin{equation}
\grad=\gs+1/\varepsilon\gf,
\end{equation}
where $\gs$ acts over the coordinates denoted by $\mathbf{x}$, and
$\gf$ acts over the fast horizontal coordinates denoted by $\fc$.

We expand the wave field $\bm{\xi}$ as a perturbative series in the
small parameter $\varepsilon$ as
\begin{eqnarray}
\bm{\xi}\left(\sc,\fc,t\right) & = & \bm{\xi_{0}}\left(\sc,\fc,t\right)+\varepsilon\bm{\xi_{1}}\left(\sc,\fc,t\right)\nonumber \\
 &  & +\varepsilon^{2}\bm{\xi_{2}}\left(\sc,\fc,t\right)+\mathcal{O}\left(\varepsilon^{3}\right).\label{eq:xi_perturbative}
\end{eqnarray}
The term at order $\varepsilon^{0}$ represents the large-scale characteristics
of the wave field, while the higher-order terms arise as the incoming
wave is scattered by the flows. This approach differs from standard
perturbative expansions in two ways --- firstly the expansion parameter
is not in terms of the flow speed $u$ but the length scale $L$;
secondly the zeroth-order field is not a solution to an equation without
flows, but senses an average effect of the flows. Our approach would
be to replace the spatially varying medium by an equivalent quiet
background, which is different from what would have been there had
there been no flows. We substitute Equation\textbf{ }(\ref{eq:xi_perturbative})
in Equation (\ref{eq:full_eqn}) and use Equation\textbf{ }(\ref{eq:material_derivative})\textbf{
}to write down the equations order by order in $\varepsilon$. 

At order $\varepsilon^{-2}$ we get 
\begin{equation}
\mathcal{L}^{\left(-2\right)}\bm{\xi_{0}}+\rho\bm{\xi_{0}}\cdot\gf\gf\phi=\bm{0},\label{eq:order_eps_m2}
\end{equation}
where 
\begin{equation}
\mathcal{L}^{\left(-2\right)}\bm{\xi}=\gf\cdot\Big\{\rho{\bf uu}\cdot\gf\bm{\xi}+\mathbf{I}\left(p-\rho c^{2}\right)\gf\cdot\bm{\xi}-p\left(\gf\bm{\xi}\right)^{T}\Big\}.\label{eq:op_ord_eps_m2}
\end{equation}
We introduce an alternate definition of the operator $\mathcal{L}^{\left(-2\right)}$
by defining the transpose map explicitly. We introduce the following
tensors and tensorial relations:
\begin{itemize}
\item The double-dot product of two dyadic tensors $\mathbf{A}$ and $\mathbf{B}$
defined as $\mathbf{A}:\mathbf{B}=A_{ij}B_{ij}$, where repeated indices
are summed over. This definition can be extended to higher order tensors
$\tilde{\mathbf{A}}$ and $\tilde{\mathbf{B}}$, where the last two
indices of $\tilde{\mathbf{A}}$ and the first two of $\tilde{\mathbf{B}}$
shall be contracted.
\item A fourth order tensor $\tilde{\mathbf{T}}$ defined as $\tilde{\mathbf{T}}_{ijkl}=\delta_{il}\delta_{jk}$.
The tensor $\tilde{\mathbf{T}}$ has the property $\tilde{\mathbf{T}}:\mathbf{D}=\mathbf{D}^{T}$
for a dyadic tensor $\mathbf{D}$. The definition can be extended
to higher-order tensors, where the transpose operator $\tilde{\mathbf{T}}$
flips the first two indices. 
\item A fourth-rank tensor $\tilde{\mathbf{I}}_{4}$, defined as $\tilde{\mathbf{I}}_{4}{}_{ijkl}=\delta_{ik}\delta_{jl}$,
that satisfies $\tilde{\mathbf{I}}_{4}:\mathbf{D}=\mathbf{D}$ for
any dyadic tensor $\mathbf{D}$.
\end{itemize}
We define the tensor\textbf{ 
\begin{equation}
\gamt=\rho{\bf uu}\cdot\tilde{\mathbf{I}}_{4}+\mathbf{I}\left(p-\rho c^{2}\right)\mathbf{I}-p\tilde{\mathbf{T}}\label{eq:gamma_tensor}
\end{equation}
}in terms of which we can rewrite the operator $\mathcal{L}^{\left(-2\right)}$
as\textbf{ 
\begin{equation}
\mathcal{L}^{\left(-2\right)}\bm{\xi}=\gf\cdot\left(\gamt:\gf\,\bm{\xi}\right).\label{eq:L-2_definition}
\end{equation}
}We shall assume henceforth that the gravitational potential $\phi$
is independent of the fast coordinate $\fc$, which lets us drop any
term like $\gf\phi$ or $\gf\gs\phi$. This assumption is reasonably
justified, since density in the near-surface layers is much smaller
than the interior and hence the gravitational potential would depend
only weakly on the flows near the surface. Under this assumption,
Eq. (\ref{eq:order_eps_m2}) becomes 
\begin{equation}
\mathcal{L}^{\left(-2\right)}\bm{\xi_{0}}=\bm{0}.\label{eq:ord_eps_m2_h}
\end{equation}
A particular solution to Equation (\ref{eq:ord_eps_m2_h}) is a zeroth-order
field $\bm{\xi_{0}}$ that is independent of the fast coordinate $\fc$.
This is a sufficient, but not a necessary condition. A general solution
to Equation (\ref{eq:ord_eps_m2_h}) can be a function of $\fc$,
which means the operator $\mathcal{L}^{\left(-2\right)}$ might be
singular. We shall need to invert this operator at order $\varepsilon^{-1}$,
so it might be necessary to define the inverse precisely by imposing
boundary conditions on the fields, or as a generalized inverse, which
is beyond the scope of this paper. We shall proceed assuming that
$\bm{\xi_{0}}$ is independent of $\fc$, which fits into our picture
of perturbations about a homogeneous background. 

Collecting terms of the order $\varepsilon^{-1}$, we obtain
\begin{equation}
\mathcal{L}^{\left(-2\right)}\bm{\xi_{1}}+\mathcal{L}^{\left(-1\right)}\bm{\xi_{0}}=\bm{0},\label{eq:order_epsm1}
\end{equation}
where 
\begin{equation}
\mathcal{L}^{\left(-1\right)}\bm{\xi}=\gf\cdot\left(\gamt:\gs\,\bm{\xi}\right),\label{eq:L-1_definition}
\end{equation}
and $\mathcal{L}^{\left(-2\right)}$ is as defined in Equation (\ref{eq:L-2_definition}).
We use the fact that $\bm{\xi_{0}}$ is independent of $\fc$ to rewrite
Equation (\ref{eq:order_epsm1}) as 
\begin{equation}
\gf\cdot\left(\gamt:\gf\bm{\xi_{1}}\right)=-\left(\gf\cdot\gamt\right):\gs\,\bm{\xi_{0}}\label{eq:order_epsm1_full}
\end{equation}
We note two facts - firstly, Equation (\ref{eq:order_epsm1_full})
is linear in $\bm{\xi_{1}}$, secondly, the right hand side is a product
of two terms, one of which varies horizontally over the slow scale
$\sc$ while the other varies over the fast scale $\fc$. This motivates
us to seek solutions for $\bm{\xi_{1}}$ by separating the slowly
varying horizontal components of $\sc$ from the rapidly varying components
of $\fc$. We substitute 
\begin{equation}
\bm{\xi_{1}}=\crctr\left(\fc,z\right):\gs\,\bm{\xi_{0}}\label{eq:xi1_xi0_rel}
\end{equation}
where $\crctr\left(\fc,z\right)$ is a third-rank tensor. Substituting
Equation (\ref{eq:xi1_xi0_rel}) in Equation (\ref{eq:order_epsm1}),
and using the fact that $\bm{\xi_{0}}$ is independent of $\fc$,
we obtain 
\begin{eqnarray}
\left[\mathcal{L}^{\left(-2\right)}\crctr+\gf\cdot\gamt\right]:\gs\,\bm{\xi_{0}} & = & \bm{0}.
\end{eqnarray}
We seek a solution $\crctr$ which satisfies 
\begin{equation}
\mathcal{L}^{\left(-2\right)}\crctr=-\gf\cdot\gamt.\label{eq:cell_problem}
\end{equation}
This is conventionally known as a cell problem \citep[see e.g.][]{bensoussan78}. 

At order $\varepsilon^{0}$ we get the equation of motion for the
zeroth-order field $\bm{\xi_{0}}$
\begin{eqnarray}
\rho\,\partial_{t}\bm{\xi_{0}}+\rho{\bf u}\cdot\partial_{t}\left(\gs\,\bm{\xi_{0}}+\gf\,\bm{\xi_{1}}\right)\nonumber \\
+\mathcal{L}^{\left(0\right)}\bm{\xi_{0}}+\mathcal{L}^{\left(1\right)}\bm{\xi_{1}}+\mathcal{L}^{\left(-1\right)}\bm{\xi_{1}}+\mathcal{L}^{\left(-2\right)}\bm{\xi_{2}}\nonumber \\
+\rho\,\bm{\xi_{0}}\cdot\gs\gs\phi & = & \bm{0},\label{eq:order0}
\end{eqnarray}
where 
\begin{eqnarray}
\mathcal{L}^{\left(0\right)}\bm{\xi} & = & \gs\cdot\left\{ \gamt:\gs\,\bm{\xi}\right\} ,\\
\mathcal{L}^{\left(1\right)}\bm{\xi} & = & \gs\cdot\left\{ \gamt:\gf\,\bm{\xi}\right\} .
\end{eqnarray}
We know that the first order field $\bm{\xi_{1}}$ depends on the
zeroth-order field $\bm{\xi_{0}}$ through Equation (\ref{eq:xi1_xi0_rel});
we substitute this in Equation (\ref{eq:order0}) and average over
the fast coordinate $\fc$. The averaging process restricts us to
the zero-spatial-frequency mode, however we do retain information
about smaller length scales through $\crctr$. Given a horizontal
cell $\mathcal{B}=[0,L)\times[0,L)$, we define the cell average of
a function $f\left(\fc\right)$ as 
\begin{equation}
\left\langle f\right\rangle =\frac{1}{L^{2}}\int_{\mathcal{B}}d\fc\,f\left(\fc\right).
\end{equation}
The terms $\mathcal{L}^{\left(-1\right)}\bm{\xi_{1}}$ and $\mathcal{L}^{\left(-2\right)}\bm{\xi_{2}}$
in Equation (\ref{eq:order0}) are total divergences in $\fc$, so
they drop off on averaging as we impose periodic boundary conditions.
We obtain
\begin{eqnarray}
\cellavg{\rho}\partial_{t}^{2}\bm{\xi_{0}}+2\cellavg{\rho{\bf u}}\cdot\gs\,\partial_{t}\,\bm{\xi_{0}}\nonumber \\
-\gs\cdot\left[\cellavg{\rho\tilde{\mathbf{C}}}:\gs\bm{\xi_{0}}\right]+\bm{\xi_{0}}\cdot\cellavg{\rho\gs\gs\phi} & = & 0,\label{eq:ord0_avg}
\end{eqnarray}
where 
\begin{equation}
\tilde{\mathbf{C}}=\gamt:\left(\tilde{\mathbf{I}}_{4}+\gf\crctr\right)\label{eq:mod_speed}
\end{equation}
resembles a wave speed squared term that governs wave propagation
on large scales. This is similar in form to the result obtained by
\citet{brown84}, except the term $\crctr$ which arises from the
separation of scales. We can substitute for $\tilde{\bm{\Gamma}}$
from Equation (\ref{eq:gamma_tensor}) to recast Equation (\ref{eq:mod_speed})
in a more tangible form as\textbf{ }
\begin{eqnarray}
\tilde{\mathbf{C}} & = & \left(c^{2}-\frac{p}{\rho}\right)\mathbf{I}\,\mathbf{I}-{\bf uu}\cdot\tilde{\mathbf{I}}_{4}+\frac{p}{\rho}\tilde{\mathbf{T}}\nonumber \\
 & + & \mathrm{terms\;arising\;from\;cell\;problem}.\label{eq:mod_speed-1}
\end{eqnarray}

Convective cells on the surface of the Sun have a structural form
where hot gas rises up near the center and cooler gas sinks down around
the edges of the cell. We assume that the flow in a particular cell
is anti-symmetric about the cell center, hence $\cellavg{\rho\mathbf{u}}=\bm{0}$.
This lets us drop the advective term in Equation (\ref{eq:ord0_avg})
and obtain 
\begin{equation}
\cellavg{\rho}\partial_{t}^{2}\bm{\xi_{0}}-\gs\cdot\left[\cellavg{\rho\tilde{\mathbf{C}}}:\gs\,\bm{\xi_{0}}\right]+\bm{\xi_{0}}\cdot\cellavg{\rho\gs\gs\phi}=\bm{0}.\label{eq:order0_simple}
\end{equation}
Equation (\ref{eq:order0_simple}) is the homogenized wave equation,
wherein we have replaced the spatially fluctuating background medium
by an effective homogeneous, anisotropic one. The effective speed
depends on both the horizontal and vertical flow profiles, the sound
speed profile and the density. If density variations in a cell are
spatially correlated with the flow speeds, a description in terms
of the average speeds and average densities which ignores correlations
might not be sufficient. 

Periodicity of flows along specific directions leads to a tensorial
wave speed, hence we expect anisotropy in wave propagation. Equation
(\ref{eq:order0_simple}) is a standard eigenvalue problem which we
can solve to find the dispersion relation $\omega\left(n,{\bf k}\right)$
for vertical order $n$ and horizontal wave vector ${\bf k}$. This
dispersion relation captures the extent to which the frequency of
a monochromatic wave would appear to differ from the canonical dispersion
relation $\omega=ck$, which is governed only by the adiabatic sound
speed. As an example, we consider waves propagating in a 2-dimensional
uniformly dense fluid. We assume that the fluid film is tiled with
small-scale, periodic flows $\mathbf{u}=\left(u_{x},u_{y}\right)$.
A plane wave propagating along the $x$ axis would have the form 
\begin{equation}
\bm{\xi_{0}}\left(\mathbf{x}\right)=\mathbf{A}\exp\left(\mathrm{i}\left(k_{x}x-\omega t\right)\right),
\end{equation}
where $\mathbf{A}=\left(A_{x},A_{y}\right)$ is the amplitude, $k_{x}$
is the $x$ component of the wave-vector and $\omega$ is the frequency
of the wave. Neglecting changes in gravitational field $\gs\phi$
in the plane, we can rewrite Equation (\ref{eq:order0_simple}) in
Fourier space as\textbf{ 
\begin{equation}
\omega^{2}\left(\begin{array}{c}
A_{x}\\
A_{y}
\end{array}\right)=k_{x}^{2}\left(\begin{array}{cc}
\cellavg{c^{2}-u_{x}^{2}+\mathrm{c.p_{1}}} & \cellavg{\mathrm{c.p_{2}}}\\
\cellavg{\mathrm{c.p_{3}}} & \cellavg{-u_{x}^{2}+\mathrm{c.p_{4}}}
\end{array}\right)\left(\begin{array}{c}
A_{x}\\
A_{y}
\end{array}\right),\label{eq:hom_fourier_dispersion}
\end{equation}
}where the terms referred to as $\mathrm{c.p}_{1}$ to $\mathrm{c.p}_{4}$
arise from the solution to the cell problem. The corresponding dispersion
relation can be obtained by diagonalizing the matrix in the second
term of Equation (\ref{eq:hom_fourier_dispersion}). If the terms
$\mathrm{c.p}_{1}$ to $\mathrm{c.p}_{4}$ are negligible, the dispersion
relation would have the form $\omega=\cellavg{\sqrt{c^{2}-u_{x}^{2}}}k_{x}$,
but this might not be a good approximation in general.

We note that Equation (\ref{eq:order0_simple}) is not damped at the
leading order, so attenuation, if any, would be captured at a higher
order. This is consistent with observations, since the observed \textit{p}-mode
line-widths are much smaller than the corresponding frequencies.

\section{\label{sec:Numerical-test}Numerical test of flow homogenization}

In this section we take a simple example of waves propagating through
flows in a 2-D plane, and show that we can effectively reproduce the
wave speed by homogenizing over small scale flows. A thorough analysis
of the approach would include realistic pressure, density and sound-speed
profiles, possibly starting from simulations of near-surface structure
of stars. In this work, we focus only on sound-speed variations induced
by flows, and assume that the background pressure and gravitational
field are unchanged in presence of the flow. We also assume that the
background medium is uniformly dense throughout. Under these assumptions,
Equation (\ref{eq:full_eqn}) simplifies considerably as we can drop
the terms $\grad\cdot\left(\mathbf{I}p\grad\cdot\bm{\xi}-p\left(\grad\bm{\xi}\right)^{T}\right)$
and $\bm{\xi}\cdot\grad\grad\phi$. We combine Equations (\ref{eq:material_derivative})
and (\ref{eq:full_eqn}) to obtain 
\begin{equation}
\partial_{t}^{2}\bm{\xi}+2{\bf u}\cdot\grad\partial_{t}\bm{\xi}+\left({\bf u}\cdot\grad\right)^{2}\bm{\xi}-\grad\left(c^{2}\grad\cdot\bm{\xi}\right)=\bm{0}.\label{eq:2d_acoustic}
\end{equation}

We assume that the waves are propagating in a medium that is periodic
along both the directions; in other words, the unit box $\mathcal{B}=[0,L)\times[0,L)$
is tessellated to fill up the entire two-dimensional plane. We choose
to work in Cartesian coordinates $\left(x,y\right)$, with the origin
at the center of a cell. As an example of a flow-field in a cell,
we choose a vortex centered about the cell-center. The functional
form of the flow is given by 
\begin{equation}
{\bf u}=u_{0}\frac{\left(x\hat{y}-y\hat{x}\right)}{\sigma}\exp\left(-\frac{\left(x^{2}+y^{2}\right)}{2\sigma^{2}}\right),\label{eq:flow_field}
\end{equation}
where $u_{0}$ is a velocity scale, $\sigma$ is a length scale of
the flow related to the cell-size $L$. The sound speed $c$ would
depend on several factors such as the pressure in a cell and the equation
of state. For our illustrative purpose, we choose a simple model given
by 
\begin{equation}
c\left(r\right)^{2}=c_{\infty}^{2}+\left(c_{0}^{2}-c_{\infty}^{2}\right)\exp\left(-\frac{x^{2}+y^{2}}{\sigma^{2}}\right),
\end{equation}
where $c_{0}$ is the sound speed at the center of the cell and $c_{\infty}$
is the sound speed far away from the center of the cell. We choose
a cell length-scale $L=1\,\mathrm{Mm}$, and values of sound speed
and flow parameters as listed in Table \ref{tab:c_u}, where $u$
refers to the magnitude of the flow velocity $\mathbf{u}$.

\begin{table*}
\begin{centering}
\begin{tabular}{|c|c|c|c|c|c|c|}
\hline 
 & $c_{0}$ & $c_{\infty}$ & $\cellavg c$ & $u_{\mathrm{max}}$ & $\sigma$ & $\cellavg u$\tabularnewline
\hline 
\hline 
Run $1$ & $10\,\mathrm{km/s}$ & $11.7\,\mathrm{km/s}$ & $11.6\,\mathrm{km/s}$ & $4\,\mathrm{km/s}$ & $0.1\,\mathrm{Mm}$ & $0.5\,\mathrm{km/s}$\tabularnewline
\hline 
Run $2$ & $10\,\mathrm{km/s}$ & $11.7\,\mathrm{km/s}$ & $11.3\,\mathrm{km/s}$ & $4\,\mathrm{km/s}$ & $0.2\,\mathrm{Mm}$ & $2.0\,\mathrm{km/s}$\tabularnewline
\hline 
Run $3$ & $10\,\mathrm{km/s}$ & $13.7\,\mathrm{km/s}$ & $12.8\,\mathrm{km/s}$ & $6.2\,\mathrm{km/s}$ & $0.2\,\mathrm{Mm}$ & $3.0\,\mathrm{km/s}$\tabularnewline
\hline 
Run $4$ & $10\,\mathrm{km/s}$ & $11.7\,\mathrm{km/s}$ & $10.7\,\mathrm{km/s}$ & $4\,\mathrm{km/s}$ & $0.4\,\mathrm{Mm}$ & $3.5\,\mathrm{km/s}$\tabularnewline
\hline 
\end{tabular}
\par\end{centering}

\protect\caption{Sound speed and flow speed corresponding to two runs with different
flow speeds in a cell. \label{tab:c_u}}
\end{table*}
We recast Equation (\ref{eq:2d_acoustic}) in the form of an eigenvalue
problem by using Equation (\ref{eq:material_derivative}) along with
Equation (\ref{eq:2d_acoustic}). Assuming harmonic time dependence,
we rewrite the system of equations as 
\begin{equation}
-\iota\omega\left(\begin{array}{c}
\bm{\xi}\\
\dot{\bm{\xi}}
\end{array}\right)=\left(\begin{array}{cc}
\mathbf{I}\left(-{\bf u}\cdot\grad\right) & \mathbf{I}\\
\mathbf{D} & \mathbf{I}\left(-{\bf u}\cdot\grad\right)
\end{array}\right)\left(\begin{array}{c}
\bm{\xi}\\
\dot{\bm{\xi}}
\end{array}\right),\label{eq:eq_mot_ham_matrix}
\end{equation}
where the operator $\mathbf{D}$ encapsulates the restoring force
on the wave field $\bm{\xi}$ and is given by $\mathbf{D}\bm{\xi}=\grad\left(c^{2}\grad\cdot\bm{\xi}\right)$.
We note that both ${\bf u}\cdot\grad$ and $\mathbf{D}$ are periodic
with the same length scale as the flow, hence we use Bloch's theorem
to reduce the problem to one cell. According to Bloch's theorem, the
wave-field $\bm{\xi}$ can be written as\textbf{ $\bm{\xi}\left(\mathbf{x};\omega\right)=\exp\left(\mathrm{i}\mathbf{k}\cdot\mathbf{x}\right)\bm{\eta}_{\mathbf{k}}\left(\mathbf{x};\omega\right)$,
}where $\mathbf{k}=\left(k_{x},k_{y}\right)$ is a horizontal, as
yet arbitrary, wave-vector, and the function $\bm{\eta}_{\mathbf{k}}$
has the periodicity of the flow. We study waves propagating along
the x-axis which have $k_{y}=0$. We impose a large-scale periodicity
condition $\bm{\xi}\left(x+2\pi R_{\odot},y;\omega\right)=\bm{\xi}\left(x,y;\omega\right)$,
which restricts us to discrete values of the wave-vector $k_{x}=\left(2\pi/R_{\odot}\right)n$
for integral $n$. Corresponding to each value of $n$ we obtain an
eigenvalue problem from Equation (\ref{eq:eq_mot_ham_matrix}), which
we solve numerically for the frequency $\omega\left(k_{x}\right)$.
We refer to this relation as `true dispersion relation'. In the limit
of infinite wavelength, we can replace the fluctuating medium by a
homogeneous one, given by Equation (\ref{eq:order0_simple}). We make
one further approximation --- because of potential non-uniqueness
in computing $\crctr$, we restrict ourselves to the zeroth-order
homogenized speed given by 
\begin{equation}
\tilde{\mathbf{C}}=c^{2}\mathbf{I}\mathbf{I}-{\bf u}{\bf u}\cdot\tilde{\mathbf{I}}_{4}.\label{eq:0th_order_hom_corr}
\end{equation}
The homogenized equivalent to Equation (\ref{eq:order0_simple}) in
this case is 
\begin{equation}
\partial_{t}^{2}\bm{\xi_{0}}-\gs\cdot\left[\cellavg{\tilde{\mathbf{C}}}:\gs\bm{\xi_{0}}\right]=\bm{0},\label{eq:2D_num_test_hom}
\end{equation}
where $\tilde{\mathbf{C}}$ is as defined in Equation (\ref{eq:0th_order_hom_corr}).
We compute the dispersion relation corresponding to Equation (\ref{eq:2D_num_test_hom}),
and refer to it as `homogenized relation'. It is worth pointing out
that the approximation made in\textbf{ }Equation\textbf{ }(\ref{eq:0th_order_hom_corr})
is crude; one can not neglect $\crctr$ if Mach numbers are close
to $1$, and it might be necessary to compute $\crctr$ by solving
Equation (\ref{eq:cell_problem}) to achieve a higher level of accuracy. 

\begin{figure*}
\includegraphics[scale=0.6]{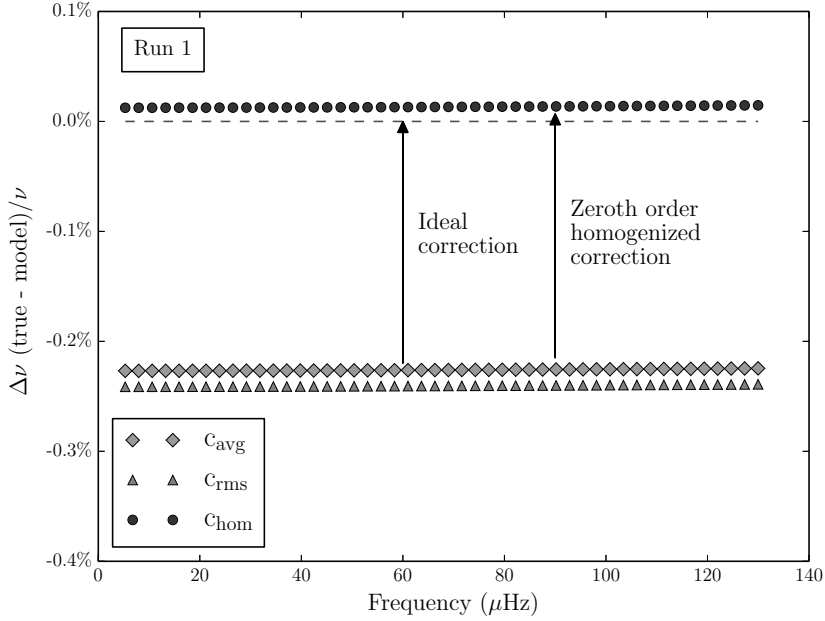}

\protect\caption{\label{fig:disp_2d_run1}Relative frequency differences between true
dispersion relation and alternate dispersion relations of the form
$\omega=\mathrm{c_{eff}}k$ for different effective speeds $\mathrm{c_{eff}}$.
Circles represent the homogenized relation (Equation \ref{eq:0th_order_hom_corr}),
diamonds represent $\omega=\protect\cellavg ck$ and triangles represent
$\omega=\sqrt{\protect\cellavg{c^{2}}}k$. We see that the homogenized
relation reproduces mode frequencies with higher accuracy than the
other relations in the low frequency limit.}
\end{figure*}

\begin{figure*}
\includegraphics[scale=0.6]{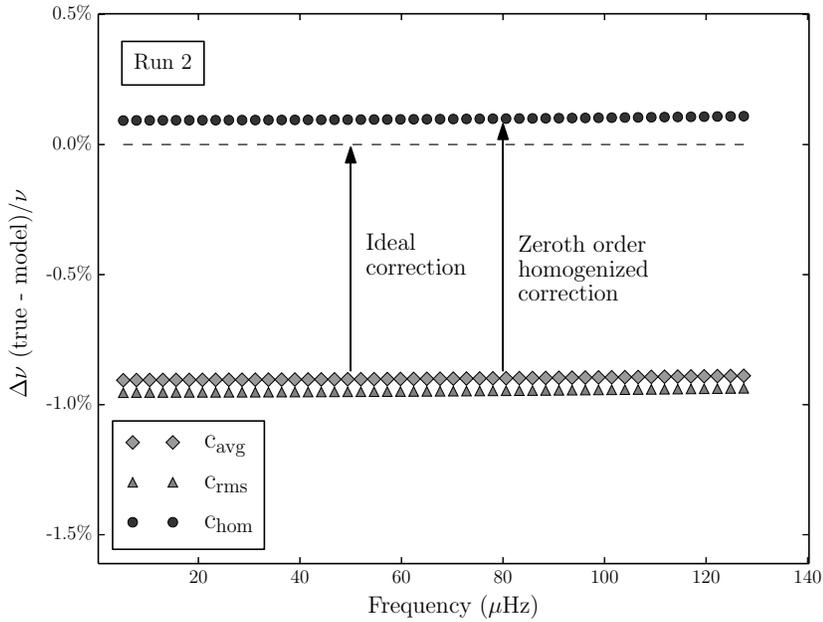}

\protect\caption{\label{fig:disp_2d_run2}Relative frequency differences as obtained
in Run $2$, where the markers represent the same quantity as Figure
\ref{fig:disp_2d_run1}. Run 2 corresponds to a higher Mach number
than Run 1, and we find that the 0-th order homogenized approximation
is not as effective at reproducing the true dispersion relation as
in Run 1. }
\end{figure*}

We compare the zeroth-order homogenized dispersion relation with the
true dispersion relation in Figure \ref{fig:disp_2d_run1}. We also
compare it with dispersion relations of the form $\omega=\cellavg ck$
and $\omega=\sqrt{\cellavg{c^{2}}}k$, which do not explicitly account
for flows. We have listed frequency differences for several different
parameter values in Table \ref{tab:Comparison-true_hom}. We find
that the homogenized relation replicates the true dispersion relation
better than the alternate relations. The relative frequency differences
corresponding to Run 2 have been plotted in Figure \ref{fig:disp_2d_run2}.
Comparing with Run 1, we note that the frequencies predicted by the
zeroth-order approximation to the homogenized dispersion reduce in
accuracy as the Mach number increases.


\begin{table*}
\begin{centering}
\begin{tabular}{|c|c|c|c|c|}
\hline 
 & $\left(\Delta\nu\right)_{\mathrm{avg}}$ (Run $1$) & $\left(\Delta\nu\right)_{\mathrm{avg}}$ (Run $2$) & $\left(\Delta\nu\right)_{\mathrm{avg}}$ (Run $3$) & $\left(\Delta\nu\right)_{\mathrm{avg}}$ (Run $4$)\tabularnewline
\hline 
\hline 
$\omega=c_{\mathrm{hom}}k$ & $0.009\,\mu\mathrm{Hz}$ & $0.07\,\mu\mathrm{Hz}$ & $-0.5\,\mu\mathrm{Hz}$ & $0.29\,\mu\mathrm{Hz}$\tabularnewline
\hline 
$\omega=\cellavg ck$ & $-0.15\,\mu\mathrm{Hz}$ & $-0.59\,\mu\mathrm{Hz}$ & $-1.8\,\mu\mathrm{Hz}$ & $-1.36\,\mu\mathrm{Hz}$\tabularnewline
\hline 
$\omega=\sqrt{\cellavg{c^{2}}}k$ & $-0.16\,\mu\mathrm{Hz}$ & $-0.62\,\mu\mathrm{Hz}$ & $-1.9\,\mu\mathrm{Hz}$ & $-1.40\,\mu\mathrm{Hz}$\tabularnewline
\hline 
\end{tabular}
\par\end{centering}

\protect\caption{Comparison between the frequency differences corresponding to the
different dispersion relation, averaged over the range $4\,\mu\mathrm{Hz}\leq\nu\leq120\,\mu\mathrm{Hz}$.
The homogenized relation, which accounts for flows explicitly, comes
closest to the true dispersion relation.\label{tab:Comparison-true_hom}}
\end{table*}

\section{Impact on solar frequencies}

Correcting the solar oscillation equations involves correcting for
both modal and model effects, as described by \citet{rosenthal_99}.
In a recent paper, \citet{piau_14} have explored the impact of various
parametrization of stellar convection --- from mixing-length theory
\citep{spruit_77} to patched hydrodynamic models --- on oscillation
frequencies. They have used horizontally-averaged sound speeds from
these models in the wave equation. We have shown that the presence
of flows introduces further advective corrections in the wave speed,
and the horizontally-averaged sound speeds do not represent the true
wave speed. The actual speed of the waves is obtained from Equation
(\ref{eq:order0_simple}), and the difference between this speed and
the horizontally-averaged sound speed can be substantial near the
solar surface. 

\begin{figure*}
\includegraphics[scale=0.8]{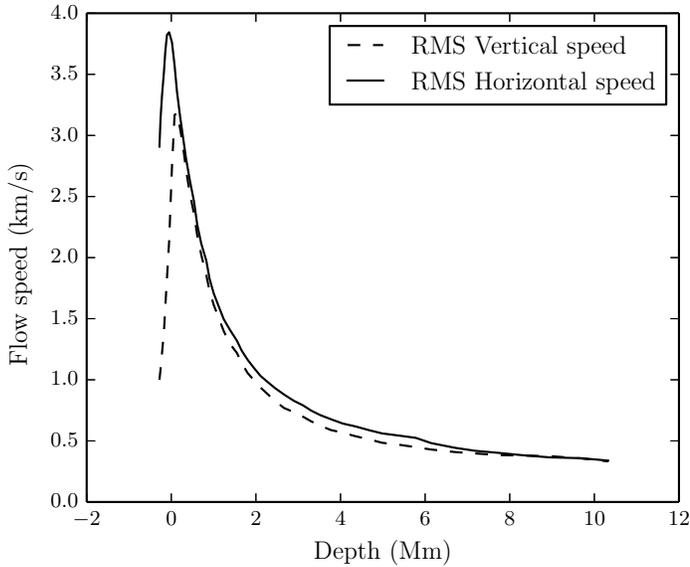}

\protect\caption{Root-mean-square flow speeds obtained from three-dimensional simulation
of the near-surface layers of the Sun \citep{voegler2005}.\label{fig:flow_profile}}

\end{figure*}

In this section, we start from Model S \citep{jcd} which uses the
mixing-length formulation of convective flux. We correct wave speeds
using flow profiles obtained from simulations by Sch\"ussler (personal
communication) using the MURaM code \citep{voegler2005}. The flow-speed
profiles have been plotted in Figure \ref{fig:flow_profile}. We start
from equation (\ref{eq:order0_simple}) and work with low-degree modes
$\left(\ell\leq3\right)$, which are predominantly radial. We assume
that the Mach number is low enough for the zeroth-order homogenization
approximation (Equation \ref{eq:0th_order_hom_corr}) to hold, and
retain only the dominant component of $\tilde{\mathbf{C}}$ which
goes roughly as $c^{2}-u^{2}$. The exact form of $u$ which goes
into the correction depends on the direction of advection at every
depth; for our purpose we shall restrict ourselves to the special
cases of radial flow $\left(\mathbf{u}={\bf u}_{\parallel}\right)$
and tangential flow $\left(\mathbf{u}={\bf u}_{\perp}\right)$. We
refer to this modified model as Model Hom, and compute the eigenfrequencies
corresponding to this model assuming the same boundary conditions
as Model S. We compare these frequencies with observed solar frequencies
obtained by the Birmingham Solar Oscillation Network (BiSON) \citep{chaplin_07}.
We find that the presence of flow advection in the wave equation leads
to a reduction in mode frequencies, which reduces the observed difference
between observed and Model S frequencies. 

\begin{figure*}
\includegraphics[scale=0.8]{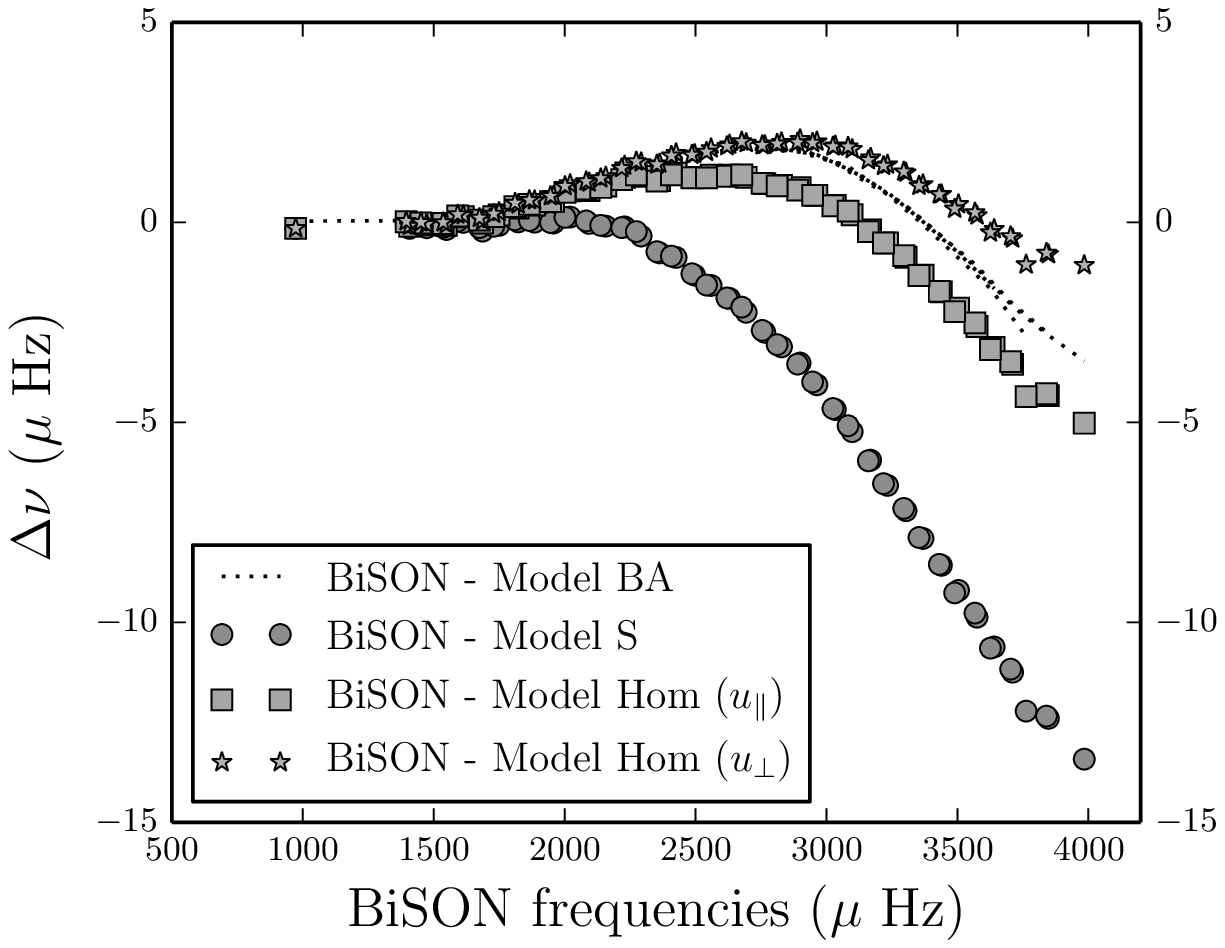}

\protect\caption{Eigenfrequencies corresponding to Model S \citep[circles]{jcd} compared
with the modified models. The squares indicate a model where advection
is predominantly radial and the stars indicate a model where advection
is predominantly tangential. The dotted line represents eigenfrequencies
corresponding to Model BA \citep{basu_antia_94}. We find that the
modified models display frequency shifts comparable to Model BA, which
shows that modal effects can be significant near the surface and need
to be taken into account.\label{fig:sol_freq}}

\end{figure*}

To study the impact of modal versus model corrections on frequencies,
we compare the Model Hom frequencies with another hydrostatic model
\citep[henceforth called Model BA]{basu_antia_94}, which uses the
Canuto-Mazzitelli formulation of convective flux \citep{canuto_mazzitelli_1991}.
The frequency differences between these models and the observations
by BiSON have been plotted in Figure \ref{fig:sol_freq}. We find
that the modal frequency shifts due to the presence of advective flows
is comparable to those obtained from modeling differences. Both the
effects need to be looked into while fitting solar frequencies.

Some of the assumptions made in this section, such as diagonally-dominated
and isotropic $\tilde{\mathbf{C}}$, are expected to be violated considerably
near the surface, but the present example demonstrates that the effect
of flow correction can be substantial. A detailed computation, starting
either from a fully hydrodynamic background or a patched hydrodynamic
background, is necessary to make quantitative statements about frequency
shifts due to convective flows.

\section{Conclusion}

Observed solar oscillation frequencies can differ from modeled frequencies
because of an inaccurate background model, as well as incomplete description
of mode physics. The presence of near-surface convective flows on
the Sun results in mode frequencies which are different from ones
computed starting from a quiet background. Three-dimensional convection
simulations generally result in an increased size of the acoustic
cavity, thereby reducing the frequencies \citep{rosenthal_99,piau_14}.
In this paper, we have shown that the impact of advection on waves
might result in frequency shifts of similar magnitudes. In presence
of flows, the appropriate wave speed is not the horizontally-averaged
sound speed, rather it is a combination of sound speed and an appropriate
projection of the flow speed in the direction of propagation.

Granulation length scales on the solar surface are much smaller than
horizontal wavelengths corresponding to low-$\ell$ modes, which lets
us replace the irregular background medium by an effective, homogeneous
one. Acoustic waves propagating through steady background flows experience
a change in phase speed on being scattered by the flows. The effective
wave equation in the large-wavelength approximation has the form 
\[
\cellavg{\rho}\partial_{t}^{2}\bm{\xi_{0}}-\grad\cdot\left[\cellavg{\rho\tilde{\mathbf{C}}}:\grad\bm{\xi_{0}}\right]+\bm{\xi_{0}}\cdot\cellavg{\rho\grad\grad\phi}=\bm{0},
\]
where the effective speed $\tilde{\mathbf{C}}$ is a tensor, given
by Equation (\ref{eq:mod_speed}).

The impact of time evolution of flows, missing in the present analysis,
needs to be studied in detail. The spatial scales of waves and background
flows are significantly different, thereby allowing a two scale analysis.
The timescales of granule evolution are, however, not very different
from time periods of the wave. Overlap of scales generally results
in strong scattering and forcing, thus a picture of flows evolving
in time might be necessary to capture frequency shifts and damping.
Added to this are the impact of super-adiabaticity and magnetic fields,
which have been ignored in the present analysis, but might be important
in governing the spectrum. Nevertheless the present analysis provides
a useful starting point to model wave propagation through convective
flows and resultant frequency shifts.

\section{Acknowledgment}

We thank the anonymous referee for his useful suggestions which improved
the quality of this work significantly. JB would like to acknowledge
the financial support provided by the Department of Atomic Energy,
India.

\bibliographystyle{apj}
\bibliography{references}

\end{document}